\documentclass[journal=jacsat,manuscript=article]{achemso}

\usepackage{chemformula} 
\usepackage[T1]{fontenc} 
\usepackage{graphicx}
\usepackage{multirow}

\author{Apurve Saini}
\affiliation[Department for Physics and Astronomy, Uppsala University, Lägerhyddsvägen 1, 752 37 Uppsala]
{Department for Physics and Astronomy, Uppsala University, Sweden}
\author{Julie A. Borchers}
\affiliation[NIST Center for Neutron Research, 100 Bureau Drive, Gaithersburg, 20899-6102, USA]
{NIST Center for Neutron Research, Gaithersburg, 20899-6102, USA}
\author{Sebastian George}
\affiliation[Department for Physics and Astronomy, Uppsala University, Lägerhyddsvägen 1, 752 37 Uppsala]
{Department for Physics and Astronomy, Uppsala University, Sweden}
\author{Brian B. Maranville}
\affiliation[NIST Center for Neutron Research, 100 Bureau Drive, Gaithersburg, 20899-6102, USA]
{NIST Center for Neutron Research, Gaithersburg, 20899-6102, USA}
\author{Kathryn L. Krycka}
\affiliation[NIST Center for Neutron Research, 100 Bureau Drive, Gaithersburg, 20899-6102, USA]
{NIST Center for Neutron Research, Gaithersburg, 20899-6102, USA}
\author{Joseph A, Dura}
\affiliation[NIST Center for Neutron Research, 100 Bureau Drive, Gaithersburg, 20899-6102, USA]
{NIST Center for Neutron Research, Gaithersburg, 20899-6102, USA}
\author{Katharina Theis-Br{\"o}hl}
\affiliation[University of Applied Sciences Bremerhaven, An der Karlstadt 8, 27568 Bremerhaven, Germany]
{University of Applied Sciences Bremerhaven, Germany}
\author{Max Wolff}
\affiliation[Department for Physics and Astronomy, Uppsala University, Lägerhyddsvägen 1, 752 37 Uppsala]
{Department for Physics and Astronomy, Uppsala University, Sweden}
\email{max.wolff@physics.uu.se}
\phone{+ 46 (0)18 - 471 3590}
\fax{+ 46 (0)18 - 471 3524}

\title[Layering of magnetic nanoparticles at amorphous magnetic templates with perpendicular anisotropy]
  {Layering of magnetic nanoparticles at amorphous magnetic templates with perpendicular anisotropy}

\abbreviations{IR,NMR,UV}

\begin{document}

\begin{tocentry}

Some journals require a graphical entry for the Table of Contents.
This should be laid out ``print ready'' so that the sizing of the
text is correct.

Inside the \texttt{tocentry} environment, the font used is Helvetica
8\,pt, as required by \emph{Journal of the American Chemical
Society}.

The surrounding frame is 9\,cm by 3.5\,cm, which is the maximum
permitted for  \emph{Journal of the American Chemical Society}
graphical table of content entries. The box will not resize if the
content is too big: instead it will overflow the edge of the box.

This box and the associated title will always be printed on a
separate page at the end of the document.

\end{tocentry}

\begin{abstract}
 We reveal the assembly of monodisperse magnetite nanoparticles of sizes 5 nm, 15 nm and 25 nm from dilute water-based ferrofluids onto an amorphous magnetic template with out-of-plane anisotropy. From neutron reflectometry experiments we extract density profiles and show that the particles self-assemble into layers at the magnetic surface. The layers are extremely stable against cleaning and rinsing of the substrate. The density of the layers is determined by and increases with the remanent magnetic moment of the particles.
\end{abstract}

\section{Introduction}
The self-assembly of colloidal particles is an attractive route for manufacturing structures with tailored mechanical \cite{Sacanna2013}, electronic \cite{RADEVA200124} or magnetic properties \cite{Yellen2012}. The equilibrium properties of soft materials exhibit a rich diversity, due to the many-body nature of the interactions (electrostatic, magnetic or steric) and deeper knowledge of colloidal systems is important for realizing smart, functional and stimuli responsive synthetic materials. Self-assembled nanostructures show remarkable collective properties \cite{D_Aguanno_1990} and are useful for engineering nanoarchitectures \cite{Kaas1983}. As an example, it has been demonstrated that the organization of nanocrystals in multi-dimensional superlattices alters their properties from their isolated counterparts \cite{Mishra_2014}. Current methodologies for assembling colloidal particles into structures include, shear \cite{solomon2006}, optics \cite{Korda2001}, depletion interactions \cite{Edwards2012, Fernandes2009}, sedimentation \cite{Lee2004, DAVIS507} as well as magnetic \cite{Yellen2012, Yellen2009} and electrical fields \cite{PhysRevE.54.496, Castellano2015}.
However, most of these methodologies face limitations as for example, colloidal particles ordered by shear may form nonequilibrium structures or optically guided assembly requires optical contrast between the medium and the particles.
The depletion interaction may lead to ordered particles that are diffusion limited with no control over orientation of the microstructure, and sedimentation typically forms irreversible structures with defects that are difficult to manipulate. Electric as well as magnetic-field-mediated colloidal self-assemblies are attractive since they can be repeatedly and reproducibly applied even in complex geometries for charged/non-charged \cite{Edwards2014, Crassous2014} and magnetic/non-magnetic particles \cite{Saini2019, Pandey2017}, respectively.\\
Directed self-assembly can be achieved by carefully choosing the building blocks and can be made very versatile by using, e.g., magnetic fields \cite{Chen2011}.
It has been shown that one-dimensional chains \cite{Tasoglu2014}, two-dimensional arrays \cite{PhysRevE.92.012303} or three-dimensional assemblies \cite{Yellen2012} can be formed.
This provides a unique route for directed self-assembly due to the instantaneous and anisotropic nature of magnetic interactions as well as its reversibility \cite{Ye2012}. Another advantage of magnetic field directed self-assembly is its non-contact nature.
The resultant magnetic colloids show huge potential for applications like medical imaging \cite{Gleich2005}, drug delivery \cite{Kenneth1978}, photonics \cite{Jianping2009}, biomedicine \cite{Kozissnik2013}, data storage \cite{Sun1989}, cellular manipulation \cite{Dobson2008}, cancer therapy \cite{Gilchrist1957, Dennis2009}, and gene transfection \cite{Scherer2002}.\\
Magnetic nano-particles (NPs) are nano-scale building blocks that follow magnetic field gradients \cite{Yellen2012}. At present, the majority of studies of the structure of magnetic fluids are devoted to bulk solutions, investigated with small angle X-ray (SAXS) \cite{Paula2019} and small angle neutron (SANS) scattering \cite{C7SM02417G}.
The self-assembly at an interface with a solid has attracted less attention but may be significantly different from that in bulk.
Even more, an interface can provide a template for the targeted self-assembly and layers may be deposited in a very controlled way.
In this context magnetically structured substrates (substrates with magnetic topographic patterns prepared on the particle scale) have been used to study the transport as well as the guided crystallization of colloidal particles.
Yellen et al. \cite{Yellen2005} used a rectangular array of cobalt microcylinders on a silicon substrate and applied a magnetic field rotating in a plane normal to the substrate.
They show that this allows the transport of non-magnetic particles dispersed in a ferrofluid.
Gunnarson et al. \cite{Gunnarsson2005} used a substrate with permalloy elliptical islands placed in a staircase-like pattern.
Applying an external magnetic field rotating in the plane of the film modifies the stray field of the magnetic ellipses and creates a driving force for the motion of paramagnetic colloidal particles placed on the film.
Tierno et al. \cite{PhysRevLett.100.148304} used magnetic garnet films, which are thin uniaxial ferromagnetic films, in which domains can be organized into symmetric patterns consisting of stripes or bubbles with perpendicular magnetization.
The resulting antiferromagnetic domains can be easily modulated in size by applying magnetic fields with a perpendicular component.
This in turn modulates the potential generated at the film surface and induces a controlled motion of paramagnetic colloidal particles placed above the film.
Particles from an aqueous solution are pinned to the Bloch wall in the film due to the intense stray field from the surface.
Applied magnetic fields can move the Bloch walls and thereby the particles.
The particles assemble into various phases depending on the complexity of domain patterns \cite{doi:10.1021/la050827c}.
The group further showed that this strategy allows separation and sorting of bi-disperse particle systems based on the particle size \cite{C5LC00067J, C6CP05599K} as well as the controlled transport of micro-sized chemical or biological cargoes by colloidal particle carriers \cite{B910427E}.\\
%
%
In complementary investigations using neutron reflectometry, we observed a close-packed wetting layer of magnetite NPs (11 nm diameter and dissolved in water) on a silicon dioxide surface\cite{C5SM00484E}.
Under an in-plane magnetic field, the particles turned and oriented with their long axis along the field direction, and under shear a static wetting layer developed directly at the surface and a depletion layer formed between the static layer and the moving ferrofluid (FF).
The self-assembly process was found to be significantly influenced by the shape anisotropy and the size distribution of the NPs. Recent polarized neutron reflectivity (PNR) studies showed that wetting and layer formation of NPs in a FF on a Si surface strongly depend on the coating, both of the substrate as well as of the particles, and can be manipulated by magnetic fields \cite{doi:10.1021/acsami.7b14849}. In this article, we investigate the self-assembly of mono-disperse magnetic NP from dilute (0.15 vol. \%) aqueous solution in the stray field of a magnetic templated substrate, a film of \ch{Tb15Co85} with out-of-plane magnetic anisotropy.
\section{Experimental}
\begin{figure}
\centering
    \includegraphics[scale=0.7]{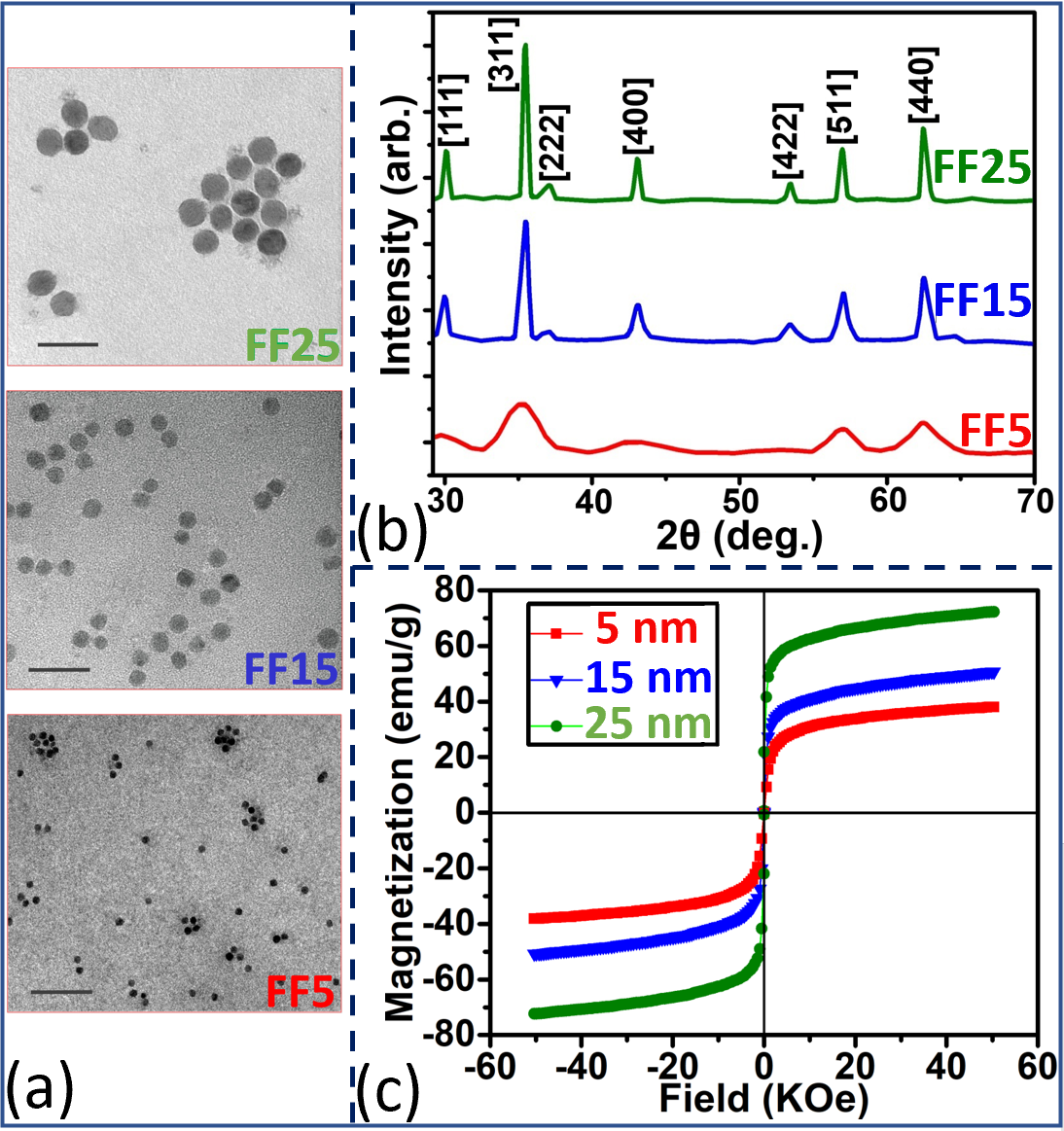}
  \caption{Panels a): TEM micrographs of iron-oxide NP samples, FF5, FF15, and FF25. Scale bars are 50 nm. Panel b): X-ray diffraction patterns of the NPs, FF5 (red), FF15 (blue), and FF25 (green) indexed according to a cubic structure. Panel c): Hysteresis loops for iron oxide nanocrystals FF5 (red), FF15 (blue) and FF25 (green) at 300 K.}
  \label{TEM_XRD}
\end{figure}

\begin{figure}
  \centering
    \includegraphics[scale=0.3]{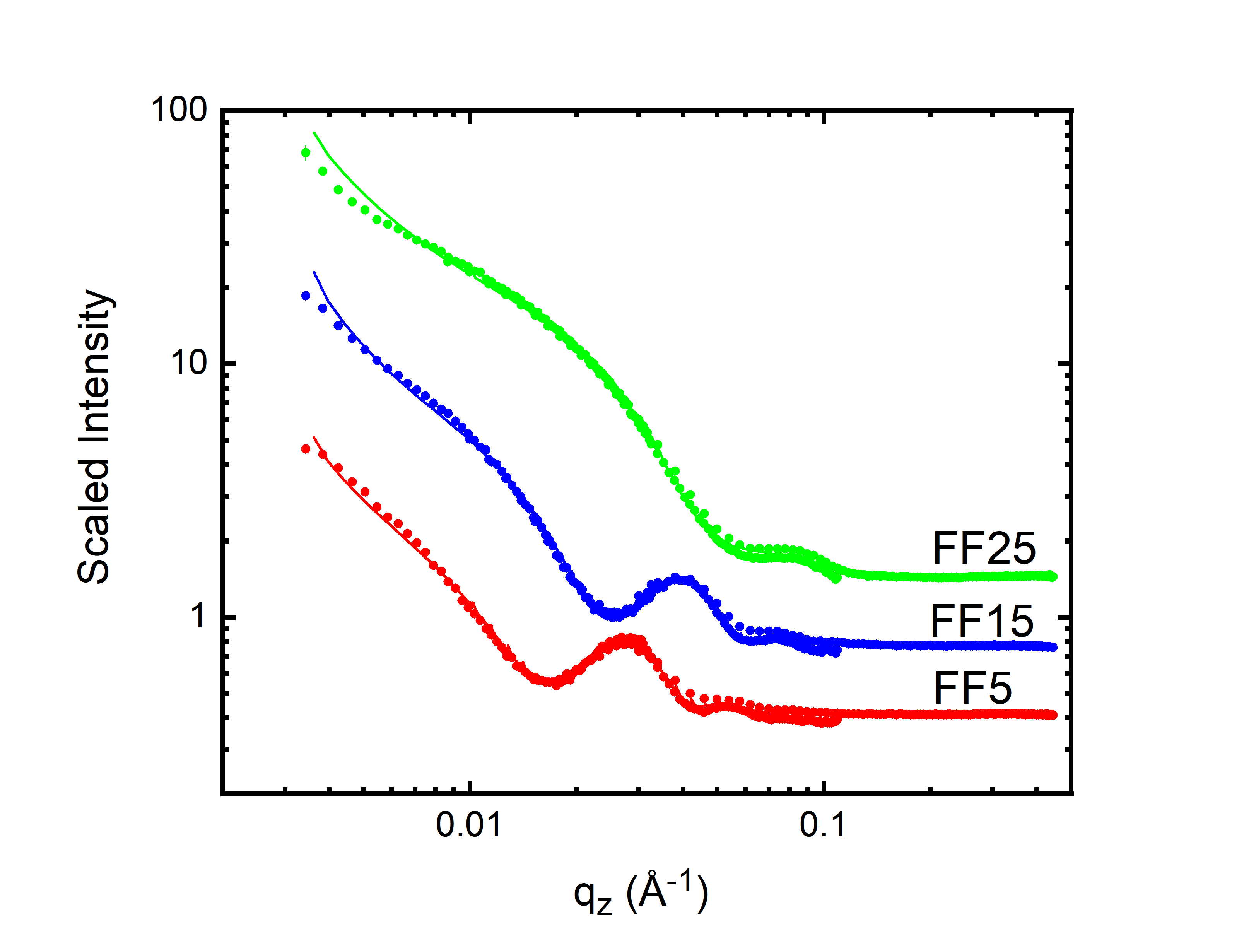}
  \caption{SANS data for samples FF5 (red symbols), FF15 (blue symbols) and FF25 (green symbols) and fits to the data (solid lines). Data for FF5 and FF15 are scaled by a factor of four and two for better visibility, respectively.}
  \label{SANS_SQUID}
\end{figure}

\begin{figure}
  \centering
    \includegraphics[scale=0.3]{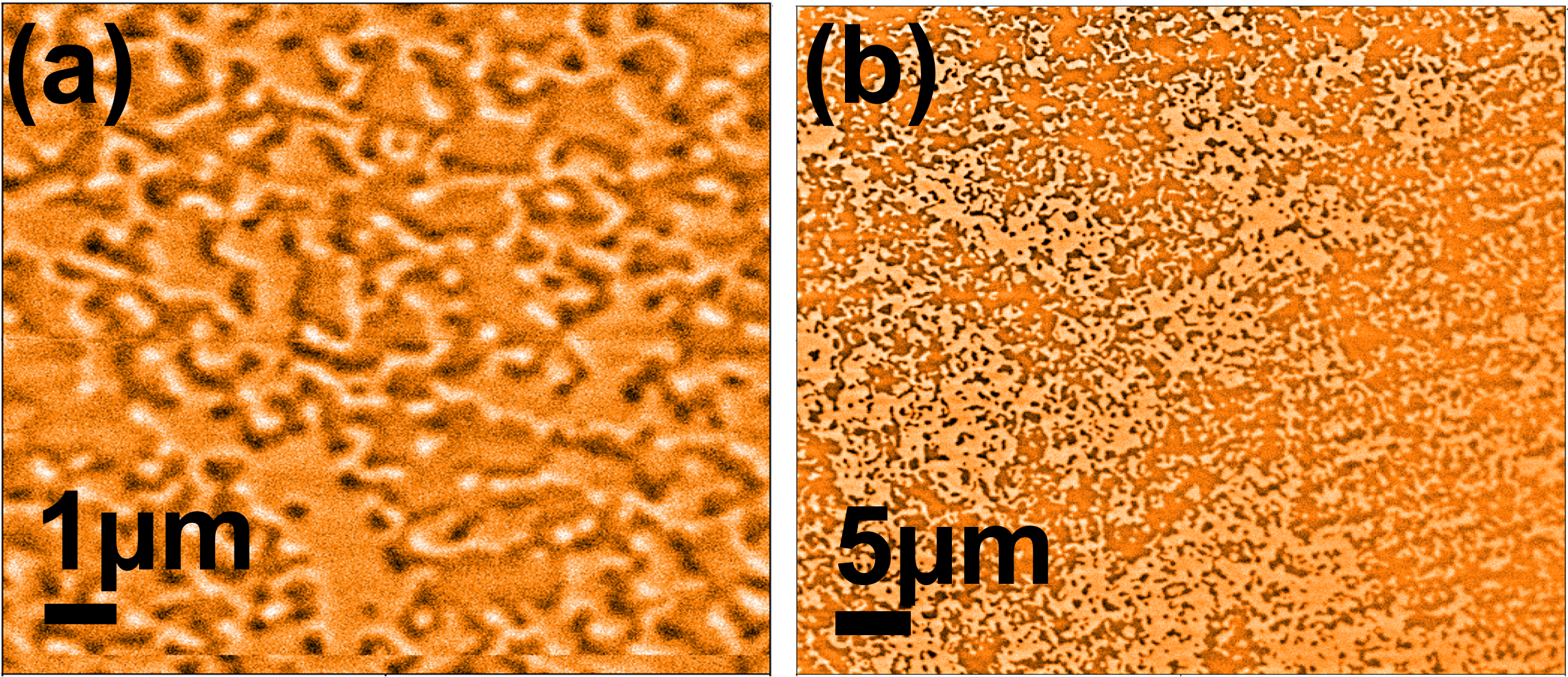}
  \caption{Domain pattern for a TbCo film, dark and bright regions correspond to areas where the sample magnetization points into or out of the sample plane, respectively.}
  \label{MFM}
\end{figure}

\begin{figure}
  \centering
    \includegraphics[scale=0.5]{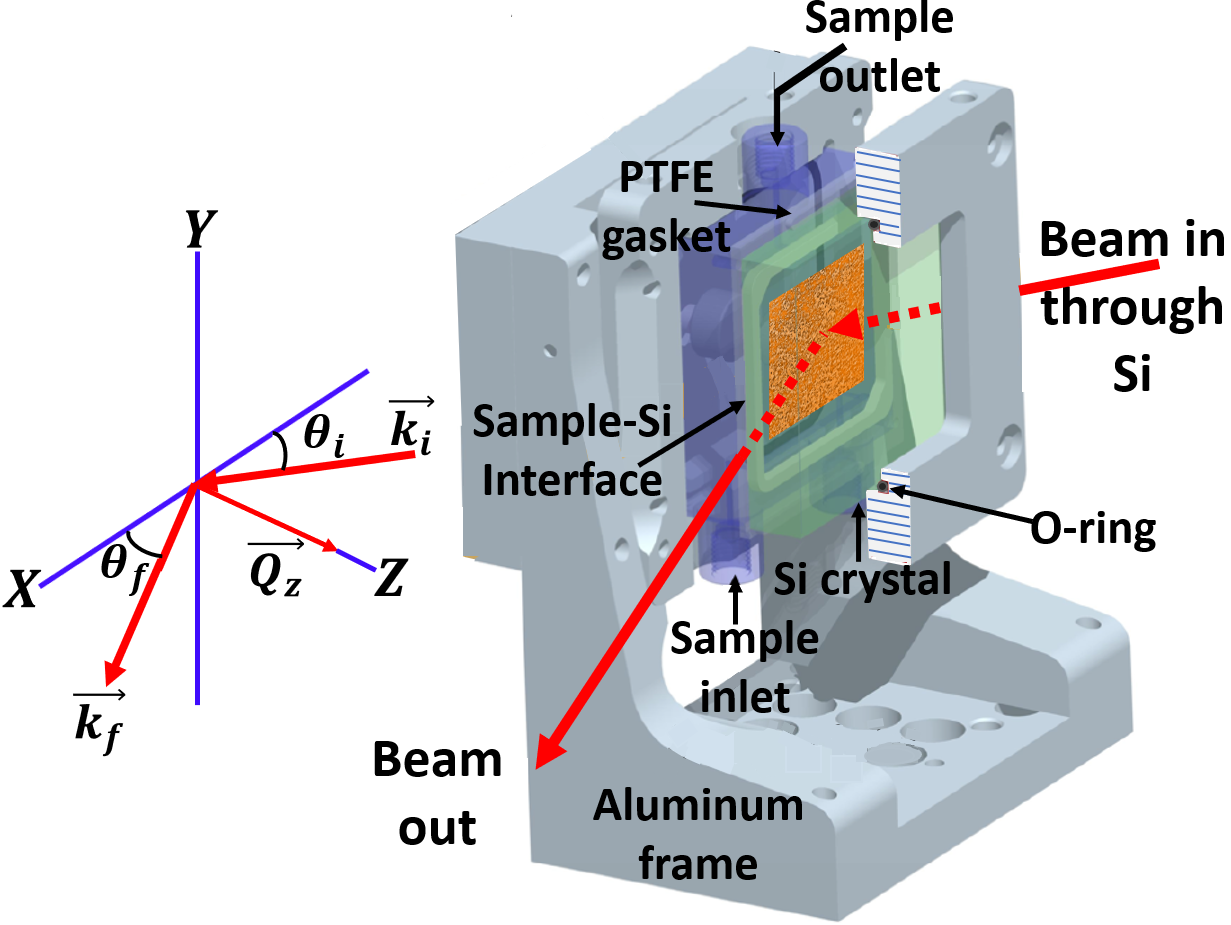}
  \caption{Sketch of the experimental setup showing the wet cell. The neutron beam and its momentum are indicated by the red arrows. The silicon crystal with the deposited magnetic template layer is shown in green. The FF sample is held inside the PTFE gasket adjacent to the magnetic template. The scattering geometry, defining incident and exit angle, is sketched at the left hand side.}
  \label{Setup}
\end{figure}

Samples FF5, FF15, and FF25 with magnetite NPs of sizes 5, 15, and 25 nm, respectively, were commercially obtained from Sigma Aldrich.\footnote{Certain commercial equipment, instruments, or materials (or suppliers, or software, ...) are identified in this paper to foster understanding. Such identification does not imply recommendation or endorsement by the National Institute of Standards and Technology, nor does it imply that the materials or equipment identified are necessarily the best available for the purpose.}
Their magnetic cores are coated with N-Hydroxysuccinimide making them very stable in water and affine to functionalized surfaces \cite{Estelrich2015}.
The size distribution and shape of the NPs were imaged with transmission electron microscopy (TEM). Representative TEM micrographs are reproduced in Fig. \ref{TEM_XRD} (panels a). The average particle diameters are 4.1 $\pm$ 0.5 nm, 14.9 $\pm$ 0.6 nm, and 22.2 $\pm$ 1.1 nm.
X-ray powder diffraction (XRD) patterns of the iron oxide nanocrystals (Fig. \ref{TEM_XRD} (panel b)) were obtained using a Philips PW 1820 diffractometer$^1$ equipped with a Cu-K$_\alpha$ X-ray source.  The mean crystal sizes of the NPs are calculated using the Scherrer equation \cite{Hall:hw0080} and are 3.9 $\pm$ 0.4 nm, 14.4 $\pm$ 0.7 nm, and 21.1 $\pm$ 1.3 nm, which are consistent with the results extracted from TEM. This shows that the NPs are single crystals.\\
Superconducting quantum interference device (SQUID) magnetometry was used to measure hysteresis loops of the powder samples at a temperature of 300 K (Fig. \ref{TEM_XRD}, panel c). The nanocrystals are superparamagnetic at room temperature with negligible coercivity. The saturation magnetization ($M_s$) is 38.0, 50.8, and 72.3 emu/g for samples FF5, FF15, and FF25, respectively. As expected $M_s$ decreases for smaller nanocrystals, due to surface spin canting and finite-size effects \cite{doi:10.1021/cm100289d, Yang2008}. All NPs have a lower saturation magnetization than bulk magnetite (92 emu/g) \cite{Wang2014}.\\
SANS measurements were done at the NGB30m SANS instrument at the NIST Center for Neutron Research (NCNR).
The NPs were diluted in a mixture of 85 \% \ch{D2O} and 15 \% \ch{H2O} , for better contract for neutrons, and contained in a titanium sample cells with quartz windows with a separation of 2 mm.
The sample-detector distances were 1, 4, and 13 m. To increase the Q-range, the detector was offset horizontally by 25 cm for the 1 m configuration.
The wavelength was $\lambda$ = 6 \AA~for all configurations and refractive neutron lenses were used for the low Q regime in the 13 m configuration.
The wavelength spread ($\frac{\Delta \lambda}{\lambda}$) was 13.8 \% (FWHM) and defined by the velocity selector in all configurations.
The data were reduced using the NCNR IGOR Pro macros \cite{Kline:do5025} with correction for scattering of the sample cell, ambient background, and flat field correction for the detector.
The collected data were normalized to the intensity of the incident beams.
A circular averaging over the detector resulted in one dimensional I(Q) curves shown in Fig. \ref{SANS_SQUID}.
The reduced SANS data were analyzed using the SasView program \cite{sasview}.
The fits, which assume a power exponent together with polydispersed core/shell spherical NPs for each sample, are tabulated in Table \ref{Result_SANS} and shown as solid lines in Fig. \ref{SANS_SQUID}. The particles have core diameters of 3.2 nm, 15 nm, and 21 nm (in line with the results from XRD and TEM), and organic shell thicknesses of 6.4 nm, 4.9 nm and 6.9 nm, respectively.
The SLD for the bulk shell material is 0.16 $\times$ 10$^{-4}$ nm$^{-2}$.
The scattering length density (SLD) of the core was fixed to 6.91 $\times$ 10$^{-4}$ nm$^{-2}$ (value for \ch{Fe3O4}). Similar to previous observation \cite{doi:10.1021/acsami.7b14849}, the SLD values for the shell material are larger than that of bulk shell material due to the presence of deuterated water in the shell, which is explained by the hydrophilicity of the head groups of the non-polar ligands attached to the NP core.\\
\begin{table}
  \caption{Results of fits to the SANS data assuming a linear combination of a power law and core/shell spheres. The SLD of the cores was fixed and the SLD of the solvent was allowed to vary in a tight range near 4.6 $\times$ $10^{-4}$nm$^{-2}$.}
  \label{Result_SANS}
  \begin{tabular}{|l|l|l|l|}
    \hline
    							& FF5		& FF15		& FF25  \\
    \hline
    Core diameter [nm]   			& 3.2(2)		& 15.4(2) 		& 21.3(2) \\
    Shell thickness [nm]			& 6.4(2) 		& 4.9(1)		& 6.9(1) \\
    Core SLD [$10^{-4}$nm$^{-2}$]	& 6.9			& 6.9			& 6.9  \\
    Shell SLD [$10^{-4}$nm$^{-2}$] 	& 2.79(10) 	& 2.40(15)		&2.94(20) \\
    Power exponent				& 1.8(1)		& 2.2(1)		& 2.3(2) \\
    Distribution radius [\%]			& 4.9			& 6.7			& 4.9\\
    Distribution shell thickness	 [\%]	& 15			& 15			& 9.1\\
    \hline
  \end{tabular}
\end{table}
Ferrimagnetic amorphous \ch{Tb15Co85} ($\approx$40 nm thick) films were grown using DC magnetron sputtering onto a piranha-etched Si crystal (50 $\times$ 50 $\times$ 10 mm) in zero magnetic field.
A 10 nm layer of amorphous \ch{Al70Zr30} was deposited as a buffer layer as well as a capping layer ($\approx$5 nm thick) to prevent oxidation \cite{KORELIS2010404}.
The resulting layer sequence was \ch{Al70Zr30/Tb15Co85/Al70Zr30/SiO2/Si}.
At this thickness the \ch{Tb15Co85} forms a worm-like domain pattern \cite{KORELIS2010404}, which was verified by magnetic force microscopy (Fig. \ref{MFM}).
For more details on the growth as well as film characteristics, we refer to literature \cite{Frisk_2015}. 
Thin films of TbCo have a strong perpendicular magnetic anisotropy \cite{Frisk_2015}. In the out-of-plane direction, the hysteresis loop is square with a large remanence, $M_r$, whereas in the in-plane direction the loop is smoothly varying with a small remanence. TbCo films are used for magnetic storage \cite{doi:10.1063/1.3703666}, spin-valve technologies \cite{doi:10.1063/1.1459605} and optical magnetic switching \cite{PhysRevLett.99.047601}.\\
For the neutron reflectometry (NR) experiments the NPs, FF5, FF15 and FF25, were dissolved in a \ch{D2O/H2O} mixture of 0.87/0.13, 0.84/0.16, and 0.77/0.23, respectively, for good contrast for neutrons, with a concentration of 0.15 vol \% \ch{Fe3O4} or 8 mg/mL. At this concentration densely packed structures close to a solid interface were reported earlier \cite{doi:10.1021/acsami.7b14849}.
The bulk SLD values of the sample components have been calculated according to ref. \cite{NCNR_SLD}. Fig. \ref{Setup} (left hand side) shows the geometry of the neutron reflectometry measurements  which were performed in a wet cell with polarization analysis.
First the substrate was measured as reference in contact with \ch{D2O}.
Then the FF-liquid with magnetic particles was measured in contact with the substrate, and finally the substrate was re-measured in contact with \ch{D2O} after thorough cleaning by three steps: First, a vigorous rinsing with pure ethanol and isopropanol, second, ultra-sonification of the crystal in a water bath for 30 minutes at 30~$^\circ$C and third, a second hard rinsing with ethanol and isopropanol followed by wiping the substrate surface with 100~\% pure fibre-made precision wipes.
Table \ref{Result_NR_sub} summarises the layer sequence of the magnetic substrate together with characteristic values extracted from the NR measurements against \ch{D2O}. These values have been kept constant for all fits with magnetic particles.\\
\begin{table}
  \caption{Magnetic template \ch{Al70Zr30/Tb15Co85/Al70Zr30} sputter grown on a Si wafer. Tabulated are fit results of NR data measured against \ch{D2O}.}
  \label{Result_NR_sub}
  \begin{tabular}{|l|l|l|l|}
    \hline
    Layer				& Thickness [nm]	& Roughness [nm]			& SLD [$10^{-4}$nm$^{-2}$]  \\
    \hline
    \ch{Al70Zr30}		& 4.96(35)			& 2.00(16)			& 1.78(23)  \\
    \ch{Tb15Co85}		& 40.47(46)		& 2.83(41)			& 2.94(13)  \\
    \ch{Al70Zr30}	 	& 11.51(27)		& 1.71(64)			&2.44(8) \\
    \ch{SiO2}			& 1.87			& 0.62(5)			& 3.8 \\
    \ch{Si}				& -				& -				& 2.07 \\ 
    \hline
  \end{tabular}
\end{table}
For the neutron reflectivity experiments, approximately 1.5 mL of each FF sample was loaded into a wet cell, which uses a silicon (111) crystal (50 $\times$ 50 $\times$ 10 mm, optically polished, obtained from CrysTec\footnote{Certain commercial equipment, instruments, or materials (or suppliers, or software, ...) are identified in this paper to foster understanding. Such identification does not imply recommendation or endorsement by the National Institute of Standards and Technology, nor does it imply that the materials or equipment identified are necessarily the best available for the purpose.}, Germany) as the reflecting interface. The FF sample was contained by a 2 mm thick polytetrafluoroethylene (PTFE) gasket between the silicon crystal and polycarbonate plate, Fig. \ref{Setup} (right hand side). During the measurements, the wet cell was oriented vertically to avoid NP sedimentation onto the Si surface due to gravity.\\
The neutron reflectivity measurements were performed on the reflectometer MAGIK at the NCNR \cite{doi:10.1063/1.2219744} with a wavelength of 5.0 \AA.
The wavelength resolution was 1.6~\% (FWHM) and the angular divergence varied from 1.4 to 1.3~\% in the investigated $Q_z$-range (both full width at half-maximum).
The collimated neutron beam penetrates the silicon crystal from the edge and undergoes reflection at the internal interfaces. The beam footprint on the sample was fixed at 25 mm.
In the case of PNR measurements (guide field at the sample position 0.7 mT) Al-coil spin flippers and Fe/Si supermirrors were used. The combined efficiency of the polarisers and spin flippers was 93~\%.
The raw data have been corrected for polarization efficiency, beam footprint and background.
No features were observed in the spin flip cross-sections and the non-spin flip channels were found to be identical.
Note, the as-grown magnetic template layer has out-of-plane anisotropy and neutrons are not sensitive to the magnetic induction along the direction of the momentum transfer.
Since, as seen from the MFM images (Figure \ref{MFM}), the magnetic domains in the substrate are randomly oriented, the signal from magnetic stray fields at the domain walls should show up in all four spin cross sections.
The absence of the signal proves that the magnetic induction resulting from stray fields averages to zero in the plane of the interface and over the coherence volume of the neutron beam independent of whether magnetic particles are present or not.
To improve the statistics we averaged both non-spin-flip channels.
The resulting data was fitted by the Parratt formalism \cite{PhysRev.95.359} implemented in the NCNR software package Refl1D \cite{Kienzle} utilizing the super-iterative algorithm \cite{doi:10.1021/cm3006887}. For each data set, the optimal number of fitting parameters (and thus the optimal number of layers) were determined using the Bayesian information criteria (BIC); $BIC = (n-\phi)\chi^2+\phi\ln (n)$, where n is the total number of data points for the measurement, $\phi$ is the total number of fitting parameters, and $\chi^2$ represents the reduced $\chi^2$ statistic of the fit, as detailed in Ref. \cite{C5SM00484E}.\\
\section{Results}
\begin{figure}
  \centering
 \includegraphics[scale=0.85]{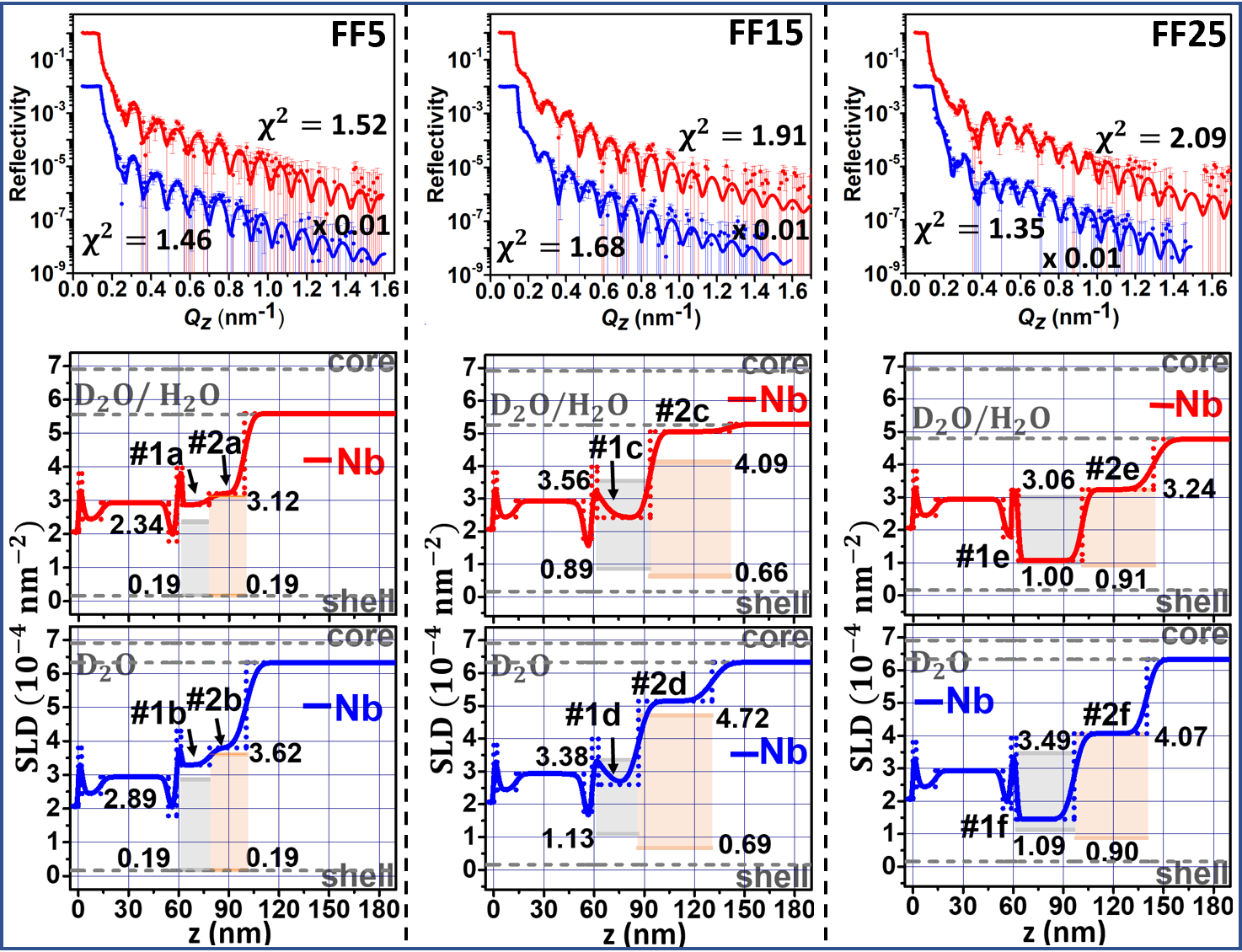}
  \caption{PNR taken for sample FF5 (a), FF15 (b) and FF25 (c) and plotted versus $Q_z$. Blue dots represent data taken with \ch{D2O} after cleaning. The solid lines represent fits to the data. The panels in the center and lower row show the resulting SLD profiles plotted as a function of distance from the Si (100) surface. For comparison, SLD values for the magnetite core, water, and shell material are indicated by grey dashed lines. The orange and yellow areas represent the SLD ranges for close-packed layers of particles with shell material (lower limits) or water (upper limits) in the inter-shell gaps.}
  \label{NR_SLD_sketch}
\end{figure}

The neutron reflectivity experiments are summarised in Fig. \ref{NR_SLD_sketch}.
In the panels on top the PNR data are plotted as a function of the wavevector $Q_z$.
The uncertainties on the individual data points correspond to $\pm$ 1 standard deviation (valid throughout the remaining text).
For all three samples two data sets are shown.
The data were taken, first, with the magnetic substrate in contact with the FF sample (red symbols) and with \ch{D2O} after cleaning (blue symbols, shifted by a factor of 100 for clarity).
The best fits to the data with the corresponding $\chi^2$ (marked), are shown as solid lines.
The SLD profiles extracted from the fits are plotted in the centre and lower rows in Fig. \ref{NR_SLD_sketch} for the measurement before and after cleaning, respectively.
The bulk SLD values of the FF components, water, shell and core material are included as grey dashed lines.
The thickness, roughness and SLD values extracted from the fits as well as the the resulting concentrations of core and shell material and water for each NP sample are summarised in Table \ref{Result_NR_sub}.\\
\begin{table}
  \caption{Parameters of the first and second particle layers at the magnetic template extracted from fits to the NR data  for FF5, FF15 and FF25, respectively.}
  \label{Result_NR_sub}
  \begin{tabular}{|c|c|c|c|c|c|c|c|}
    \hline
   Sample 				&Layer 	& Thickness	& Roughness 	& SLD				& \multicolumn{3}{c|}{Composition [\%]}	\\
   					&		& [nm]		& [nm]		&  [$10^{-4}$nm$^{-2}$]	& Core	& Shell	& \ch{D2O}	\\
    \hline
    \multirow{2}{*}{FF5}	&1a   	& 16.0(3)		& 0.48(27)		& 2.87(5)  				& 0.50	&  50.1	& 49.4		\\
   					&2a		& 21.2(3)		& 4.5(3)		& 3.21(7)				& 0.38	& 43.8	& 55.8		\\
    \hline
   FF5				&1b		& 17.2(3)		& 0.96(23)		& 3.29(4)				& 0.46	& 49.3	& 50.2		\\
   after cleaning			&2b		& 21.9(4)		& 4.0(4) 		& 3.8	1(3)				& 0.37	& 40.9	& 58.8		\\
    \hline
   \multirow{2}{*}{FF15}	&1c   	& 31.9(3)		& 7.5(3)		& 2.4	2(5)				& 10.8	& 59.3	& 29.9		\\
   					&2c		& 45.5(4)		& 4.1(2)		& 5.06(2)				& 7.4		& 6.7		& 85.9		\\
    \hline
   FF15				&1d		& 24.0(4)		& 7.7(3)		& 2.61(3)				& 14.4	& 61.6	& 24.0		\\
   after cleaning			&2d		& 44.0(6)		& 5.1	(3)		& 5.1	5(2)				& 7.8		& 19.9	& 72.3		\\
    \hline
   \multirow{2}{*}{FF25}	&1e   	& 38.2(2)		& 1.0	(2)		& 1.1	0(3)				& 12.5	& 86.0	& 1.5			\\
   					&2e		& 42.8(3)		& 3.3(2)		& 3.24(3)				& 11.1	& 38.5	& 50.4		\\
    \hline
   FF25				&1f		& 34.7(4)		& 1.2(2)		& 1.45(4)				& 13.7	& 80.4	& 5.9			\\
   after cleaning			&2f		& 43.6(5)		& 4.5	(2)		& 4.07(2)				& 10.9	& 37.7	& 51.4		\\
    \hline
  \end{tabular}
\end{table}
The fitting model describes two NP layers adjacent to the magnetic substrate.
These layers consist of core and shell material as well as water.
Our main question is whether a layer is built of densely packed NPs.
The dense packing is defined by the SLD calculated from the fractional packing of the NPs.
In the model arrangement, the particles are closely packed in a hexagonal close packed (CP) arrangement to form a 2D sheet (see supplementary material), with the core/shell diameter being the lattice parameter for the unit cell.
For the calculation we assume hard spheres with a particle-to-particle separation distance equal to the sum of the core diameter and twice the shell thickness.
From these assumptions the volume fraction filled by core and shell material can be calculated (see supplementary material).
Since the spheres do not fill the space completely, voids remain.
These may be filled either by ligands (shell material) or deuterated water.
The presence of excess shell material has been reported earlier \cite{C5SM00484E}.
Since the SLD of the shell material (hydrogenated) and the deuterated water differ significantly, a wide range in SLDs can describe CP layers but with different relative concentrations of water and ligands.
The range of SLDs defined in this way is indicated by the orange and yellow areas marked in the SLD profiles for the first and second interfacial layer of particles, respectively.
Layers may be CP if the SLD falls within the orange or yellow areas and loose packed (LP) if it is larger than the maximum SLD value of the range. 
Note, firstly, that with increasing core diameter the relative volume of the cores with respect to the rest of the layers increases and the range of SLDs defining CP layering shrinks.
At the same time the CP region shifts towards slightly larger SLDs, since the SLD of the cores is much larger than that of the ligands.
Secondly, the CP range defined as described above depends on the thickness of the layers, which may slightly vary as a result of the deformation of the NPs shells in the layers.
This is in particular true for the second layer of particles, where the thickness exceeds one particle diameter.
Due to the roughness between the layers and the limited Q-range of the NR data, we were not able to resolve sublayers, dividing each of the NP layers into two layers of mainly shell material and a center layer, that is more rich in cores, as proposed previously \cite{C5SM00484E, doi:10.1021/acsami.7b14849}.
For the details of the calculations we refer to the supporting information.\\
\section{Discussion}
\begin{figure}
  \centering
    \includegraphics[scale=0.5]{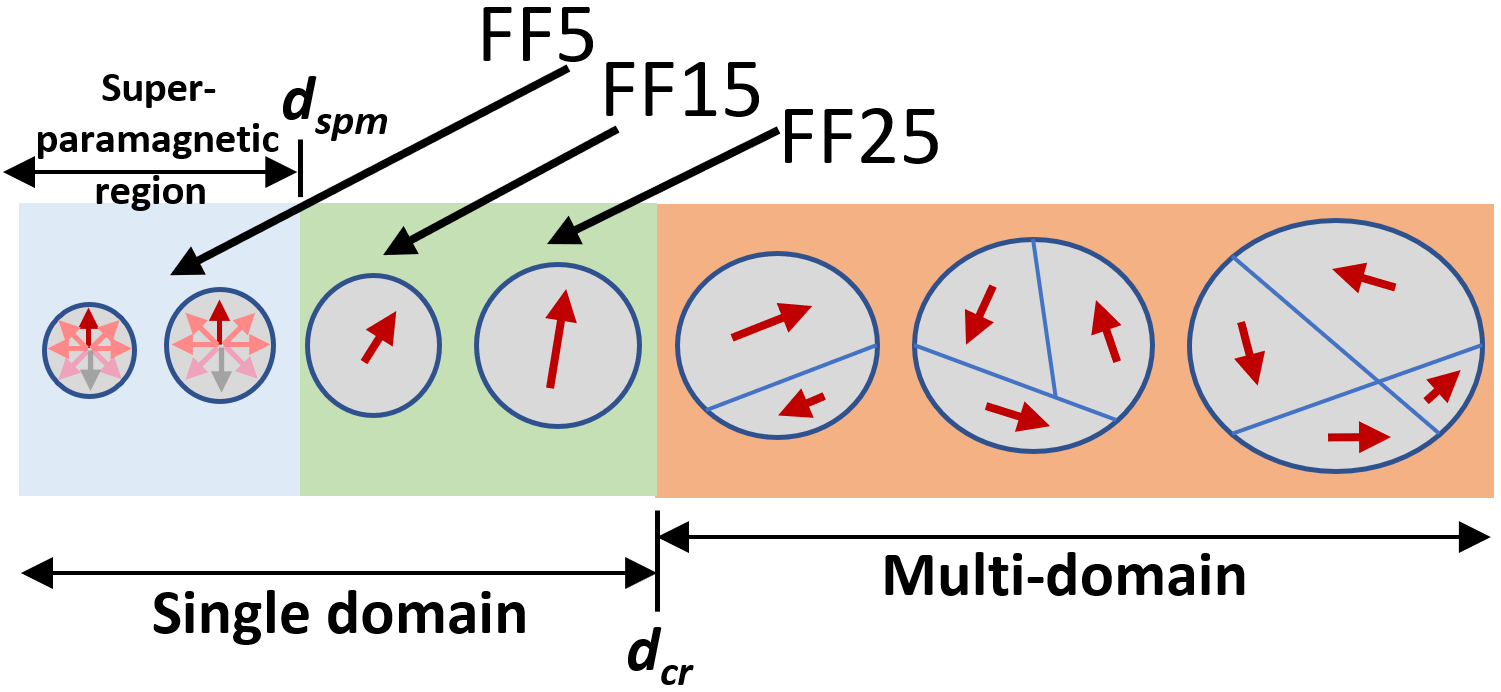}
  \caption{Schematics of particle magnetisation for different particle sizes.}
  \label{Particle_mag}
\end{figure}

The magnetic properties of NPs can be categorised with respect to their size \cite{6172415}, see schematic in Fig. \ref{Particle_mag}.
Iron oxide-based NPs with sizes above a critical diameter, $d_{cr}$, for the formation of domains are multi domain and display coercivity as well as remanent magnetization \cite{Scherer2002}.
The parameter $d_{cr}$ depends on material and is around 30 nm for single crystalline \ch{Fe3O4} \cite{Odenbach}.
The magnetization reversal takes place via domain wall motion.
Below $d_{cr}$, the formation of domains is not energetically favorable and the particles are single domain \cite{C4SM01308E}.
For such particles the magnetization of the whole particle must rotate into the field direction and away from the anisotropy direction K, which costs the energy $E = KV$ (V = volume) \cite{Petinov2014}.
As a consequence, single domain particles have a larger coercive field compared to multi-domain particles.
If dissolved in a liquid, the magnetization can reorient by rotation of the particle itself into the field direction (Brownian relaxation), the superparamagnetism becomes connected to Brownian motion rather than magnetic anisotropy.
This process is energetically more favorable than the Néel relaxation. For particles of diameter D < $d_{spm}$ (superparamagnetic limit), the anisotropy energy KV becomes comparable to thermal energy $k_{B}T$ \cite{C3SM00132F}. As a result, the particle becomes magnetically unstable and spontaneously changes the magnetization direction (Néel relaxation) and is intrinsically superparamagnetic \cite{doi:10.1063/1.3665886}.
For \ch{Fe3O4} $d_{spm}$ is below 15 nm \cite{6172415}. In summary, the NPs in sample FF5 are magnetically unstable and intrinsically superparamagnetic, while the ones in FF15 and FF25 are single domain and ferromagnetic but with a larger magnetization per particle in FF25. Fig. \ref{Particle_mag} summarises the magnetic state of particles of increasing size. The sizes of the NPs investigated in this work are marked as well.\\
To discuss the results extracted from the fits, we first focus on the layer thicknesses extracted for each sample and summarised in Table \ref{Result_NR_sub}.
From Table \ref{Result_SANS} we can calculate the diameter of the NPs including the shell and get approx. 16 nm, 25 nm and 35 nm for FF5, FF15 and FF25, respectively.
Comparing these numbers to the thicknesses of the first particle layer of approx. 16 nm, 32 nm and 38 nm extracted from the NR experiments shows that we can identify a particle wetting layer at the interface with the topmost \ch{Al_70Zr_30} layer for all the samples.
Interestingly, this situation does not change after cleaning and rinsing of the surfaces and re-measuring them in contact with \ch{D2O}.
The wetting layer is still present, however, with a slightly reduced thickness and increased SLD, which is explained by the collapse of the shells during the cleaning procedure and rehydration in \ch{D2O} during the measurement, respectively.
The exception found for sample FF5 is explained by the fact that for this sample the first and second interfacial layer have very similar SLDs and can hardly be separated.
For the second layer we find thicknesses of 21 nm, 46 nm and 43 nm, which exceed the diameter of the particles.
Moreover, from the SLD profiles shown in Fig. \ref{NR_SLD_sketch}, it is seen that between the second layer and the bulk FF, a smooth transition region exists.
Both these observations are in line with a second layer, that is still well assembled but subject to large fluctuations with respect to roughness, resulting from the increased distance from the magnetic template.
Similar to the first layer we find this layer to be stable against the cleaning and rinsing procedure but this time with almost no changes in SLD, thickness and roughness.\\
The self-assembly of particles in FF5, FF15 and FF25 is defined by dipole interactions, resulting in a force on the magnetic particle in a magnetic field gradient.
Such a field gradient arises from the stray fields at the domain walls of the magnetic template layer and the particles will move towards higher magnetic flux densities, which are largest right at the surface of the magnetic template layer.
As a consequence it is expected that larger NPs self-assemble better than smaller ones.
Indeed, the SLD value extracted for sample FF5 is larger than the SLD range calculated for the full coverage with CP particles (Fig. \ref{NR_SLD_sketch}) and is explained by the presence of water close to the interface.
For FF15 and FF25 the SLD values are consistent with a full coverage with CP particles, however, with water present in the layer for sample FF15 and only core and shell material for FF25 (Table \ref{Result_NR_sub}).
For the second layer only the SLD values extracted for the largest particles, FF25, from the fits (Table \ref{Result_NR_sub}) can be explained by a CP but hydrated layer.
The particles in sample FF25 are the largest particles investigated but still single domain and have the highest magnetisation and strongest interaction \cite{doi:10.1002/smll.201101456}.\\
As explained above for the three samples, only minute changes in reflectivities are found after thorough cleaning.
The dipolar forces are strong enough to stabilize the self assembled structures.
To further confirm this, Si substrates coated with TbCo and buffer layers were immersed into the FF samples for 5 hrs.
Then SEM micrographs were taken after following the cleaning procedure identical to the one used during the neutron studies.
\begin{figure}
  \centering
    \includegraphics[scale=0.5]{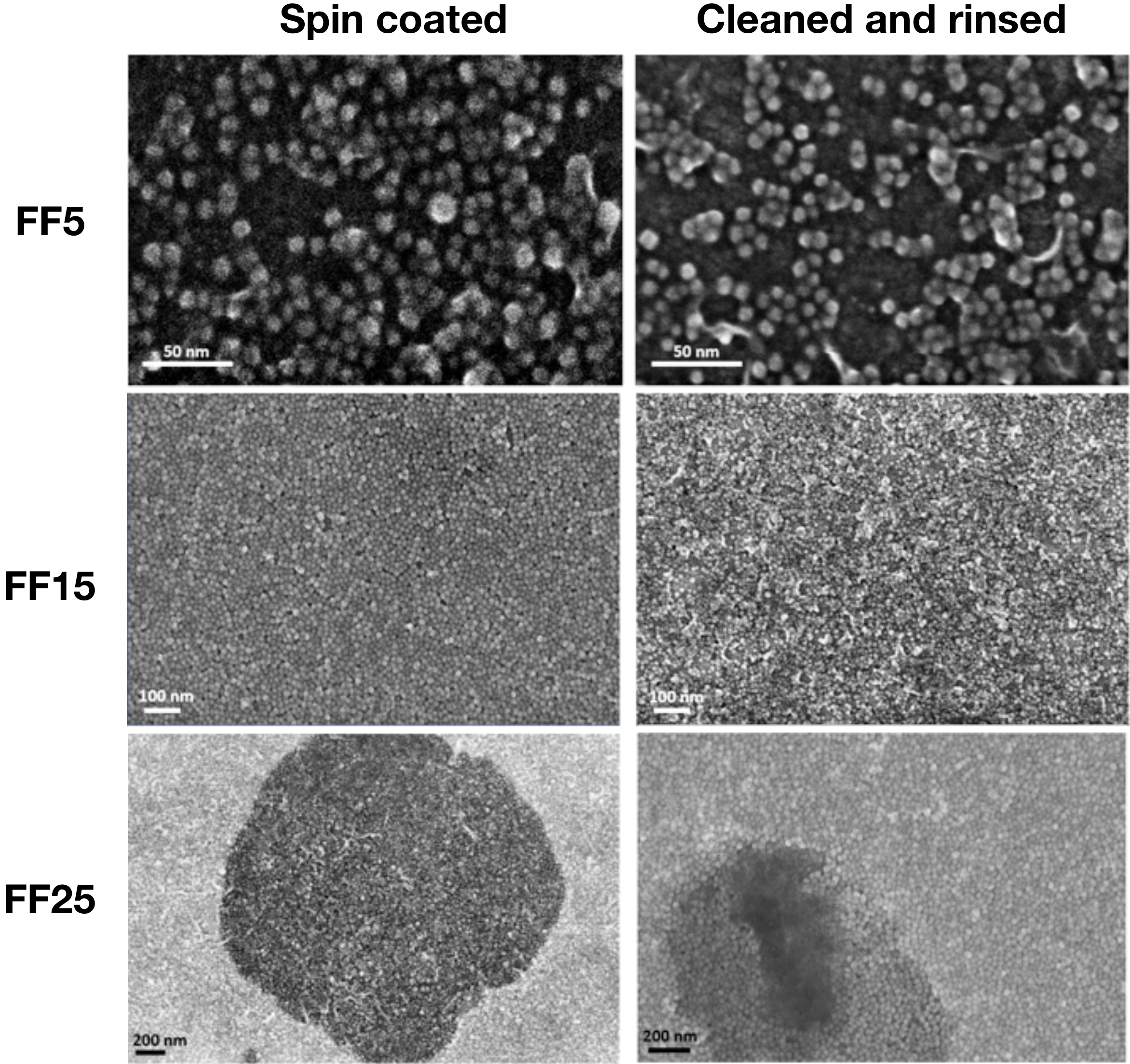}
  \caption{The upper, middle and lower panels show SEM images of sample FF5, FF15 and FF25 deposited onto TbCo, respectively. The panel to the left and right were taken before and after cleaning, respectively.}
  \label{SEM}
\end{figure}
The left and right panels of Fig. \ref{SEM} show SEM images for samples FF5 (top panels) and FF15 (middle panels) and FF25 (lower panels) before and after cleaning.
For FF5, the NPs are clearly visible in both images.
It turns out that after the cleaning procedure the density of particles remains almost unchanged and patches of CP particles are visible together with areas of bare substrate.
A similar effect is found for the NPs of sample FF15 but with a perceivable larger surface coverage.
For sample FF25 (Fig. \ref{SEM}), before and after the cleaning, both layers of NP can be distinguished in the SEM images and can be perceived as CP.
In contrast, applying the same cleaning procedure to FF25 particles chemically bound to (3-Aminopropyl) triethoxysilane (APTES) coated Si wafer (see Supporting information) through amide linkages, results in removal of more than 95 \% of the particles (for details see Supporting information).\\
The SEM images described above are in qualitative agreement with the SLD profiles extracted from the NR experiments.
However, NR averages over the coherence volume of the beam and the SLDs tabulated in Table \ref{Result_NR_sub} correspond to average values over several tens of $\mu$m (larger than the size of the SEM images), while SEM is very sensitive to local defects, providing complementary information.
The SEM images for sample FF5 show patches of adsorbed NPs rather than a continuous wetting layer.
This indicates that the NR measurements might have to be interpreted in this way.
Instead of assuming a CP or LP layer, we may assume CP patches with gaps, filled by water in between.
Indeed the stray field from the magnetic template is strongest at the domain walls and it is expected that the particles settle preferentially there \cite{doi:10.1021/la050827c}.\\
The NPs in FF25 form a CP layer with low SLD, which is explained by assuming only ligands filling the voids between particles.
Since the same chemical interactions are likely to hold in the cases with islands of CP particles we can assume that the CP patches are also composed of core and shell material for the other layers and samples and then calculated the minimum surface coverage, $SC_{min}$, of CP patches.
\begin{equation}
SC_{min} = \frac{SLD_{Sol.}-SLD_{meas.}}{SLD_{Sol.}-SLD_{Lig.dens.}}
\end{equation}
Here, $SLD_{Sol.}$, $SLD_{meas.}$ and $SLD_{Lig.dens.}$ is the SLD of the solution, the measured SLD and the SLD calculated assuming only ligands in the voids of the layer, respectively.
Table \ref{Surf_Cov} summarises the results of these calculations.
\begin{table}
  \caption{Surface coverage of the substrate with CP patches of NPs assuming only ligands in the voids. Indices 1 and 2 indicate the first and second wetting layer, respectively.}
  \label{Surf_Cov}
  \begin{tabular}{|l|l|l|l|}
    \hline
    Sample								& FF5	& FF15	& FF25  \\
    \hline
    $SLD_{Sol.}$ [$10^{-4}$nm$^{-2}$]		& 5.4		& 5.2		& 4.8  \\
    $SLD_{meas.1}$ [$10^{-4}$nm$^{-2}$]		& 2.9		& 2.4		& 1.1  \\
    $SLD_{meas.2}$ [$10^{-4}$nm$^{-2}$]		& 3.3		& 2.6		& 1.5 \\
    $SLD_{Lig.dens.1}$ [$10^{-4}$nm$^{-2}$]	& 0.19	& 0.89	& 1.0 \\
    $SLD_{Lig.dens.2}$ [$10^{-4}$nm$^{-2}$]	& 0.19	& 0.66	& 0.91 \\ 
    $SC_{min1}$	 [\%]						& 48		& 65		& 97\\
    $SC_{min2}$	 [\%]						& 41		& 57		& 86\\
    \hline
  \end{tabular}
\end{table}
The calculated coverage is in qualitative agreement with the SEM images.
Note, any solvent present in the patches will increase the surface coverage so a slightly higher surface coverage seen in the SEM images is expected.
\section{Conclusion}
In conclusion, we report magnetic self-assembly of superparamagnetic (5 nm) and single domain ferromagnetic spherical iron-oxide NPs (15 nm and 25 nm) from dilute (0.15 vol. \%) aqueous solution onto a ferrimagnetic substrate (\ch{Tb15Co85}) with perpendicular (out-of-plane) anisotropy.
We find the particle size and resulting magnetic moment as key factor for the formation of dense layers.
Larger NPs (approx. 25 nm size particles), which are single domain and have a comparatively large moment, show the most pronounced layering at the solid surfaces.
For the smallest (5 nm) particles, the dominance of Brownian motion over Neel relaxation results in less pronounced layering.
The self-assembled layers for all samples are firmly attached and stable even after a thorough cleaning of the substrate.
A comparison of NR and SEM results shows that patchy areas of dense layers are formed, which are probably pinned at the domain walls.\\
%

\begin{acknowledgement}

Access to the NGB SANS was provided by the Center for High Resolution Neutron Scattering, a partnership between the National Institute of Standards and Technology and the National Science Foundation under Agreement No. DMR-1508249.  The authors thank Cedric Gagnon for assistance with the instrumentation and Gabriella Andersson for fruitful discussions.

\end{acknowledgement}

\begin{suppinfo}

The supporting information contain the model calculations of the SLDs assuming voids being filled with ligands or solvent for all layers.
The resulting values are tabulated and SEM images taken for a sample of the same NP concentration but in contact with an APTES substrate before and after cleaning are shown.

\end{suppinfo}

\bibliography{References}

\end{document}